\documentclass[aps,reprint,longbibliography,showkeys,superscriptaddress]{revtex4-2}
\usepackage{bm,graphicx,tabularx,array,booktabs,dcolumn,xcolor,microtype,multirow,amscd,amsmath,amssymb,amsfonts,physics,tikz,wrapfig,enumitem,float,upgreek}

\usepackage[version=4]{mhchem}
\usepackage[utf8]{inputenc}
\usepackage[T1]{fontenc}
\usepackage[normalem]{ulem}

\usepackage{}
\usepackage[]{graphicx}

\usepackage{siunitx}
\usepackage{fixltx2e}
\DeclareSIUnit\hartree{E\textsubscript{h}}

\usepackage{hyperref}
\hypersetup{
    colorlinks=true,
    linkcolor=blue,
    filecolor=blue,      
    urlcolor=blue,	
    citecolor=blue
}

\usepackage{titlesec}

\usepackage{caption} 
\usepackage{subcaption}
\usepackage{cleveref}

\newcommand{\hU}{\hat{U}}

\newcommand{\hkap}{\hat{\kappa}}

\usepackage{mathtools}

\newcommand{\hE}{\hat{E}}

\newcommand{\UCAM}{Yusuf Hamied Department of Chemistry, University of Cambridge, Lensfield Road, Cambridge, CB2 1EW, U.K.}

\begin{document}

\title{Accurate and gate-efficient quantum ans\"atze for electronic states without adaptive optimisation}
\author{Hugh~G.~A.~Burton}
\email{hgaburton@gmail.com}
\affiliation{\UCAM}
\date{\today}

\begin{abstract}
\normalsize
The ability of quantum computers to overcome the exponential memory scaling of many-body problems is 
expected to transform quantum chemistry.
Quantum algorithms require accurate representations of electronic states on a quantum device, but
current approximations struggle to combine chemical accuracy and gate-efficiency while preserving physical symmetries,
and rely on measurement-intensive adaptive methods that tailor the wave function \textit{ansatz} to each molecule.
In this contribution,
we present a symmetry-preserving and gate-efficient \textit{ansatz}
that provides chemically accurate molecular energies with a well-defined 
 circuit structure.
Our approach exploits local qubit connectivity, orbital optimisation, and connections with 
generalised valence bond theory to maximise the accuracy that is obtained with shallow quantum circuits.
Numerical simulations for molecules with weak and strong electron correlation, 
including benzene, water, and the singlet-triplet gap in tetramethyleneethane, 
demonstrate that chemically accurate energies are achieved with as much as 84\,\% fewer
two-qubit gates compared to state-of-the-art adaptive \textit{ansatz} techniques.
\end{abstract}

\maketitle

\section{Introduction}
Solving the electronic Schr\"{o}dinger equation underpins theoretical predictions of
chemistry.
Since exact solutions formally scale exponentially with the number of electrons, we currently rely on 
polynomially-scaling approximations, such as coupled-cluster and density functional theory. 
However, these methods fail when electronic states cannot be easily approximated, 
due to strong spin-coupling or competing electronic configurations.
Gate-based quantum computation promises to solve these strongly correlated 
problems by representing electronic states using polynomially scaling 
quantum resources.\cite{McArdle2020}
For near-term quantum hardware, which is limited to shallow circuits,
the most promising methods optimise a parametrised \textit{ansatz} 
for the electronic state using algorithms such as the variational quantum eigensolver (VQE).\cite{Peruzzo2014}
However, traditional wave function approximations cannot be easily translated into quantum 
circuits, and no consensus has been reached on the best \textit{ansatz} 
for practical quantum computation.\cite{Anand2022,Tilley2022} 

We define a quantum \textit{ansatz} as a parametrised
unitary transformation $\hU$ applied to an initial qubit state $\ket{\Phi_0}$, where $\hU$ is composed of building blocks 
$\hU = \hU_1(\theta_1) \cdots \hU_M(\theta_M)$ that are implemented using quantum gate operations.
This \textit{ansatz} should (i) be highly accurate and systematically improvable, (ii)
correspond to a shallow quantum circuit, and (iii) satisfy the physical symmetries of the Hamiltonian
including the particle number, Pauli antisymmetry, and the $\expval*{\hat{S}^{2}}$ and
 $\expval*{\hat{S}_z}$ spin quantum numbers.
Generally, $\hat{U}$ is constructed using either 
hardware-efficient operators, which give shallow circuits but fail to preserve physical 
symmetries,\cite{Kandala2017,Ryabinkin2018,DCunha2023,Cerezo2021}
or fermionic operators that preserve physical symmetries, but require a large number of gate operations.\cite{Anand2022,Lee2019c,Sokolov2020,Filip2020,Matsuzawa2020,Kottmann2022,Kottmann2023,Ryabinkin2018a,Xia2021} 
For example, methods based on unitary coupled cluster (UCC) theory  expand the 
exponential of a sum of non-commuting  fermionic  operators using a Trotter approximation
$\mathrm{e}^{\hat{A}+\hat{B}} = \lim_{m\rightarrow \infty} (\mathrm{e}^{\hat{A}/m}\mathrm{e}^{\hat{B}/m})^m$,
which is only exact in the infinite limit.\cite{Anand2022,Peruzzo2014,Grimsley2020}
Overcoming this trade-off between gate efficiency and physical symmetries is a major challenge to achieve
chemically meaningful calculations on near-term devices.

The disentangled UCC approach
showed that exact states can be represented using an infinite product of 
exponential one- and two-body fermionic operators.\cite{Evangelista2019}
However, the accuracy of truncated expansions strongly depends on the choice and 
ordering of operators.\cite{Evangelista2019,Grimsley2020,Izmaylov2020}
Therefore, state-of-the-art algorithms 
use an adaptive \textit{ansatz} to select the most relevant operators from a pre-defined pool.
For example, the ADAPT-VQE protocol  adds operators with the largest energy improvement on each 
iteration,\cite{Grimsley2019,Tang2021,Chan2021,Tsuchimochi2022,Yordanov2021}
while  DISCO-VQE performs a coupled global optimisation of the continuous operator amplitudes
and discrete operator sequence.\cite{Burton2023}
These methods have shown that accurate and gate-efficient approximations can 
be identified, while even shallower  circuits can be constructed using qubit-excitation-based 
operators that ignore fermionic antisymmetry.\cite{Yordanov2021,Magoulas2023a}
However, adaptive optimisation has high quantum measurement costs that are likely to
preclude  simulations on real hardware, and can yield very different circuits with inconsistent
hardware noise  along potential energy surfaces.

In this work, we show that adaptive optimisation can be avoided using a well-defined \textit{ansatz}
structure.
We present the tiled Unitary Product State (tUPS) approximation and show  that it can 
provide high-accuracy, spin-preserving, and gate-efficient quantum circuits. 
Building on the quantum number preserving (QNP) gate fabric,
which combines symmetry-preservation with local qubit connectivity,\cite{Anselmetti2021} 
our approach maximises the accuracy for shallow quantum circuits by incorporating
orbital optimisation and an initial qubit state motivated by perfect-pairing valence bond theory.
We demonstrate that chemical accuracy (within $1.59\,\mathrm{mE_h}$) 
can be obtained for potential energy surfaces and spin-state energies in molecules
with weak and strong electron correlation, using up to 84\,\% fewer two-qubit gates compared to state-of-the-art adaptive algorithms.
Our results comprehensively show that the fixed UPS \textit{ansatz} can exceed the accuracy and gate
efficiency of adaptive optimisation methods, paving the way for a new generation
of electronic structure approximations for quantum algorithms.

\section{Theory}

\subsection{Tiled unitary product states}
\label{sec:tups}
It has recently been shown that an arbitrary wave function
can be constructed from a  product of $M$ unitary fermionic operators as\cite{Evangelista2019}
\begin{equation}
\ket{\Psi(\bm{\theta})} = \prod_{I=1}^{M} \exp(\theta_I \hkap_{\mu_I})\ket{\Phi_0},
\label{eq:ups}
\end{equation}
where $\hkap_{\mu_I}$ correspond to generalised one- and two-body fermionic operators acting between arbitrary orbitals, which may appear multiple times.
Ref.~\onlinecite{Burton2023} showed that Eq.~\eqref{eq:ups} can be restricted to include
only spin-adapted one-body operators $\hkap^{(1)}_{pq}$
or paired two-body operators $\hkap^{(2)}_{pq}$ acting between spatial orbitals $\phi_p$ and $\phi_q$, defined as
\begin{equation}
\hkap^{(1)}_{pq} = \hE_{pq} - \hE_{qp},
\quad\text{and}\quad
\hkap^{(2)}_{pq} = \hE_{pq}^2 - \hE_{qp}^2.
\label{eq:opss}
\end{equation}
Here, $\hE_{pq} = \hat{p}^{\dagger} \hat{q} + \hat{\bar{p}}^{\dagger} \hat{\bar{q}}$ 
is the singlet excitation operator,\cite{HelgakerBook}  $\hat{p}^{\dagger} $ ($\hat{\bar{p}}^{\dagger} $) 
defines the creation operator for a high (low) spin electron in spatial orbital $\phi_p$,
$\ket{\Phi_0}$ is the initial state, and $\theta_I$ are continuous parameters.
Each operator may appear multiple times with a different continuous parameter,
and the particular sequence of operators is indexed by the discrete variables $\mu_I$.
Any state with the same particle number and spin symmetry as $\ket{\Phi_0}$ 
can be represented for $M\rightarrow \infty$ using a suitable operator sequence and continuous variables.
Following Ref.~\cite{Burton2023}, we refer to the \textit{ansatz}  in
Eq.~\eqref{eq:ups} as a Unitary Product State (UPS) to highlight its general  mathematical structure.

The operators in Eq.~\eqref{eq:opss} are generalised excitations that are not restricted to occupied-virtual transitions.
The $k$-UpCCGSD approach proposed in Ref.~\cite{Lee2019c} was the first quantum \textit{ansatz} to use only these one-body and paired two-body generalised fermionic operators, showing that the gate count can be reduced to linear scaling with the number of spin orbitals.

\begin{figure*}[t]
\centering
\includegraphics[width=0.95\linewidth]{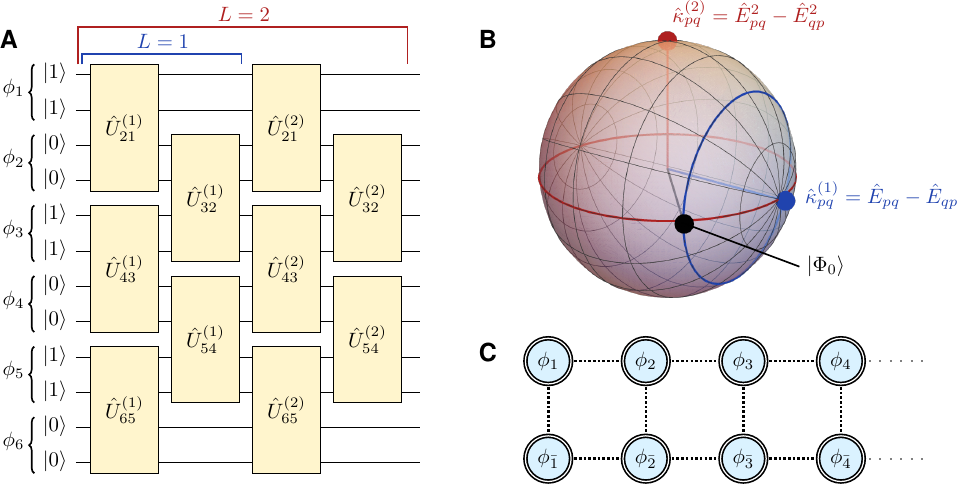}
\caption{\small (\textbf{\textsf{A}}) The tUPS \textit{ansatz} employs a tiled gate fabric with $L$ layers containing
operators that only couple adjacent spatial orbitals.
(\textbf{\textsf{B}}) Spin-adapted one-body and paired-two body operators correspond to Givens rotations 
around orthogonal axes in the singlet subspace where the spatial orbitals $\phi_p$ and $\phi_q$ contain a total of 
one high-spin and one low-spin electron. Arbitrary states within this subspace exist on the surface of a
sphere due to normalisation.
(\textbf{\textsf{C}}) Ordering the qubits such that the high-spin orbitals appear before the low-spin orbitals
removes all the Pauli-$Z$ strings from the Jordan--Wigner encoding of $\hkap_{pq}^{(1)}$ and  $\hkap_{pq}^{(2)}$,
and a suitable 2D arrangement means that only local connectivity between qubits with the same spatial orbital is required.
Note that this ordering is not shown in the circuit diagram (\textbf{\textsf{A}}).}
\label{fig:tups}
\end{figure*}

In practice, we require a truncated UPS with a finite number of  operators and a shallow circuit implementation. 
Building on the QNP gate fabric,\cite{Anselmetti2021}
we define the tiled Unitary Product State (tUPS) using $L$ layers of fermionic operators 
that act between sequential spatial orbitals (Fig.~\ref{fig:tups}\textcolor{blue}{A}) as
\begin{equation}
\ket{\Psi_{\text{tUPS}}} = \prod_{m=1}^{L} 
\qty(
\prod_{p=1}^{A} \hat{U}^{(m)}_{2p+1,2p}
\prod_{p=1}^{B} \hat{U}^{(m)}_{2p,2p-1}  
)\ket{\Phi_0},
\label{eq:tups}
\end{equation}
where $A=\frac{N-2}{2}$ or $\frac{N-1}{2}$ and $B=\frac{N}{2}$ or $\frac{N-1}{2}$  for an even or odd number of spatial orbitals $N$.
The operators $\hat{U}^{(m)}_{pq}$ each contain three unique variable parameters for every layer $m$ and are constructed from
a paired two-body operator sandwiched between two spin-adapted one-body operators  as 
\begin{equation}
\hat{U}^{(m)}_{pq} = \exp(\theta^{(m)}_{pq,1}\, \hkap_{pq}^{(1)}) \exp(\theta^{(m)}_{pq,2}\, \hkap_{pq}^{(2)}) \exp(\theta^{(m)}_{pq,3}\, \hkap_{pq}^{(1)}).
\label{eq:ugate}
\end{equation}
So far, this circuit is almost identical to the QNP approach, but our definition of  $\hat{U}^{(m)}_{pq}$ 
contains two one-body operators rather than just one, 
giving faster convergence with respect to the number of layers. 
The accuracy of the tUPS approximation can be increased further by
including orbital optimisation and modifying the initial qubit register using connections
to the perfect pairing  valence bond theory,\cite{Lawler2010, Beran2005,Voorhis2000,Hurley1953,Lehtola2018}
leading to the orbital-optimised (oo-tUPS) and perfect-pairing (pp-tUPS) variants of the tUPS approximation, respectively (\textit{vide infra}).

\subsection{Optimal ordering of unitary operations}
\label{sec:liealgebra}
The favourable properties of $\hU_{pq}^{(m)}$can be understood
by characterising
how the exponential operators  $\mathrm{exp}(\theta \, \hkap_{pq}^{(1)})$ and  
$\mathrm{exp}(\theta \, \hkap_{pq}^{(2)})$ transform the electronic Hilbert space.
For a $\mathcal{D}$-dimensional Hilbert space and $\theta \in \mathbb{R}$, 
the exponential operators $\mathrm{exp}(\theta\, \hkap_{I})$
belong to the $\mathrm{SO}(\mathcal{D})$ matrix group 
and the anti-Hermitian operators $\hkap_{I}$ belong to the associated 
Lie algebra $\mathfrak{so}(\mathcal{D})$.\cite{HallBook,GilmoreLieBook}
Therefore, the operators $\mathrm{exp}(\theta \, \hkap_{pq}^{(1)})$ and  $\mathrm{exp}(\theta \, \hkap_{pq}^{(2)})$ are isomorphic to rotations in Euclidean space. 
These rotations can be illustrated by considering the action of $\hkap^{(1)}_{pq}$ and 
$\hkap^{(2)}_{pq}$ on the singlet states for two electrons in two spatial orbitals
$\phi_p$ and $\phi_q$, for which $\mathcal{D}=3$.
The complete $\mathfrak{so}(3)$ Lie algebra is then characterised by the commutation relations
\begin{subequations}
\begin{align}
[\hkap^{(1)}_{pq},\hkap^{(2)}_{pq}] &= 2\hkap^{(3)}_{pq},
\\
[\hkap^{(2)}_{pq},\hkap^{(3)}_{pq}] &= 2\hkap^{(1)}_{pq},
\\
[\hkap^{(3)}_{pq},\hkap^{(1)}_{pq}] &= 2\hkap^{(2)}_{pq},
\end{align}
\end{subequations}
where $\hkap^{(3)}_{pq} \equiv 2 [\hkap^{(1)}_{pq},\hkap^{(2)}_{pq}]$ by definition.
Matrix representations for these operators in the two-electron singlet basis 
$\{\ket{q\bar{q}} , \frac{1}{\sqrt{2}}\qty(\ket{q\bar{p}} + \ket{p\bar{q}} ) , \ket{p\bar{p}}    \}$, 
where e.g.\ $\ket{q\bar{q}} = \hat{\bar{q}}^{\dagger}\hat{q}^{\dagger}\ket{\text{vac}}$, are given by
	\begin{subequations}
		\begin{align}
		\hat{k}^{(1)}_{pq}  &= 
		\begin{pmatrix} 0 & -\sqrt{2} & 0 \\ 
       \sqrt{2} & 0 & -\sqrt{2} \\ 
        0 & \sqrt{2} & 0 \end{pmatrix},
		\\
		\hat{k}^{(2)}_{pq}  &= 
		\begin{pmatrix} 0 & 0 & -2 \\ 
        0 & 0 & 0 \\ 
        2 & 0 & 0 \end{pmatrix},
	    \\ 
		\hat{k}^{(3)}_{pq}  &= 
		\begin{pmatrix} 0 & \sqrt{2} & 0 \\ 
        -\sqrt{2} & 0 & - \sqrt{2} \\ 
        0 & \sqrt{2} & 0 \end{pmatrix}.
	\end{align}
	\end{subequations}
Therefore, these operators are isomorphic to generators of rotations on the singlet hypersphere\cite{Burton2022}
(Fig.~\ref{fig:tups}\textcolor{blue}{B}) through e.g.\ 
$	\hkap^{(1)}_{pq} \cong \hat{L}_x$, $\hkap^{(2)}_{pq} \cong \hat{L}_y$, and $\hkap^{(3)}_{pq} \cong \hat{L}_z$.
For many-electron systems, $\hat{k}_{pq}^{(1)}$ and $\hat{k}_{pq}^{(2)}$ 
mix determinants with  
the same electronic occupation excluding $\phi_p$ and $\phi_q$, 
and form a universal set of Givens rotations,\cite{Arrazola2022}
as  described in Appendix~\ref{sec:givens}.

In 3D Euclidean space, any rotation $\hat{R}$ of the axis system can be parametrised using three Euler angles
$(\theta_1, \theta_2, \theta_3)$ and two rotation axes (e.g.\ $x$ and $y$) as\cite{GilmoreLieBook} 
\begin{equation}
\hat{R}(\theta_1, \theta_2, \theta_3) = \exp(\theta_1  \hat{L}_x) \exp(\theta_2 \hat{L}_y) \exp(\theta_3 \hat{L}_x).
\label{eq:EulerAngles}
\end{equation}
Therefore, 
the singlet Hamiltonian for two-electrons in $\phi_p$ and $\phi_q$ can be completely
diagonalised using the $\hat{U}^{(m)}_{pq}(\theta_1, \theta_2, \theta_3)$ operation
proposed in Eq.~\eqref{eq:ugate}, and we do not need to explicitly consider $\hkap^{(3)}_{pq}$.
In contrast, the operators
$\hat{U}^{\text{QNP}}_{pq}(\theta_1, \theta_2)  = \mathrm{exp}(\theta_1\, \hkap^{(1)}_{pq}) \mathrm{exp}(\theta_2\, \hkap^{(2)}_{pq})$
used in the QNP \textit{ansatz}\cite{Anselmetti2021} 
cannot perform an arbitrary axis transformation in this 3-dimensional Hilbert space.\cite{Evangelista2019,Izmaylov2020}

The ability of $\hat{U}^{(m)}_{pq}(\theta_1, \theta_2, \theta_3)$ to perform any transformation 
within the two-electron, two-orbital Hilbert space provides the tUPS \textit{ansatz} with 
greater variational flexibility than the QNP gate fabric, 
particularly when the wave function is not dominated by a single determinant.
Therefore, the tUPS \textit{ansatz} will converge faster than the QNP approach 
with respect to the number of layers  
for deep quantum circuits and multi-configurational ground states.
Furthermore, we can exploit a four-point and eight-point parameter shift rule\cite{Wierichs2022} 
to  evaluate analytic partial derivatives with respect to the tUPS parameters,
as detailed Appendix~\ref{sec:shift}.

\subsection{Convergence to the exact ground state}
\label{sec:exact}
Convergence to the exact ground state for $L \rightarrow \infty$ is guaranteed 
by the relationship of the tUPS \textit{ansatz} to the general UPS structure [Eq.~\eqref{eq:ups}] and stems
from disentangling the fermionic Lie algebra.\cite{Evangelista2019,Izmaylov2020}
Since every many-body excitation can be expressed using nested 
commutators 
containing only operators of the form $\hkap^{(1)}_{pq}$ and $\hkap^{(2)}_{pq}$, 
the spin-adapted one-body and paired two-body operators can ``generate'' the full 
Lie algebra of many-body excitations.\cite{Evangelista2019,Burton2023}
Expanding the product of  exponential operators using the Baker--Campbell--Hausdorff
theorem,\cite{HallBook} 
$\mathrm{e}^{\hat{A}}\mathrm{e}^{\hat{B}} = \mathrm{e}^{\hat{A}+\hat{B} + \frac{1}{2}[\hat{A},\hat{B}] + \cdots}$,
ensures that Eq.~\eqref{eq:ups} can
represent any fermionic unitary transformation for a sufficiently large $L$.
This result extends to the tUPS approach by noting that 
any spin-adapted one-body or paired two-body operator can be represented as a sum 
of operators and commutators that only include  $\hkap^{(1)}_{p,p\pm1}$ or $\hkap^{(2)}_{p,p\pm1}$, 
as shown in Appendix~\ref{sec:universal}.

Conserving quantum numbers such as the particle number, $\expval*{\hat{S}_z}$, 
and  $\expval*{\hat{S}^2}$ ensures that approximate wave functions
can be physically interpreted and have a high overlap with the true ground state, making 
them suitable initial states for fault-tolerant quantum algorithms.\cite{McArdle2020,SLee2023}
Employing fermionic operators means that truncated tUPS approximations 
conserve particle number and Pauli antisymmetry, which can be broken using 
gate-efficient qubit excitation operators.\cite{Yordanov2021,Magoulas2023a,Magoulas2023b,Xia2021,Ryabinkin2018a}
The $\expval*{\hat{S}_z}$
and  $\expval*{\hat{S}^2}$ expectation values  of  $\ket{\Phi_0}$ are conserved 
because the spin-adapted one-body and paired two-body operators commute with the 
spin operators, e.g.\ $[\hat{S}_z, \hkap^{(1)}_{pq}] = 0$ and $[\hat{S}^2, \hkap^{(1)}_{pq}] = 0$.
Furthermore, these operators can be implemented exactly as a quantum circuit,
whereas spin-adapted unpaired two-body operators contain a sum of non-commuting terms that requires a
Trotter approximation, which destroys the spin adaptation.\cite{Tsuchimochi2020}

\subsection{Efficiency of quantum resources}
\label{sec:gates}
In practice, the fermionic operators must be expressed as 
elementary qubit gates using transformations such as 
the Jordan--Wigner (JW) encoding\cite{Jordan1928} 
\begin{subequations}
\begin{align}
\hat{p}^{\vphantom{\dagger}} &= \frac{1}{2}(\hat{X}_{p} + \mathrm{i} \hat{Y}_p) \prod_{r=0}^{p-1} \hat{Z}_r,
\\
\hat{p}^{\dagger} &= \frac{1}{2}(\hat{X}_{p} - \mathrm{i} \hat{Y}_p)\prod_{r=0}^{p-1} \hat{Z}_r.
\end{align}
\label{eq:jw}
\end{subequations}
Here, $\{\hat{X}_p, \hat{Y}_p, \hat{Z}_p \}$ are the Pauli operators for the $p^\text{th}$ qubit
and  $\prod_{r=0}^{p-1} \hat{Z}_r$ encodes the
fermionic anti-symmetry. 
Since two-qubit CNOT gates create more noise than single-qubit gates, 
the CNOT count is commonly used to assess the practicality of a quantum circuit.
For arbitrary many-body excitation operators, the CNOT count 
 is dominated by the Pauli-$Z$ string and increases
with the number of orbitals.\cite{Yordanov2020}
This cost can be reduced by using the qubit creation and 
annihilation operators 
$\hat{Q}^{\dagger}_p = \frac{1}{2}(\hat{X}_{p} - \mathrm{i} \hat{Y}_p)$ and 
$\hat{Q}^{\vphantom{\dagger}}_p = \frac{1}{2}(\hat{X}_{p} + \mathrm{i} \hat{Y}_p)$ to define
qubit-excitation-based (QEB) operators.\cite{Yordanov2020,Yordanov2021,Yordanov2022}
While one- and two-body QEB operators yield efficient circuit 
implementations,\cite{Yordanov2020} and can be used in ADAPT-VQE,\cite{Yordanov2021,Yordanov2022}
they ignore Pauli antisymmetry and can 
destroy the sign structure of the wave function.

The tUPS \textit{ansatz} avoids any compromise between CNOT efficiency and symmetry 
preservation.
If the spin-orbitals are indexed such that $\hkap^{(1)}_{p,p\pm1}$ acts between adjacent qubits,
with the ordering
\begin{equation}
\{ \phi_1, \phi_2, \dots, \phi_N,  \phi_{\bar{1}},   \phi_{\bar{2}}, \dots, \phi_{\bar{N}} \},
\label{eq:order}
\end{equation}
then the JW encoding of the spin-adapted one-body  operator becomes equivalent to a 
QEB single excitation and the corresponding unitary operation becomes
\begin{equation}
\begin{split}
&\exp(\theta\, \hkap^{(1)}_{p,p\pm1}) = 
\exp(\mathrm{i} \frac{\theta}{2} (\hat{X}_{p} \hat{Y}_{p\pm1} - \hat{Y}_{p} \hat{X}_{p\pm1} ))
\\
&\quad\quad\quad\times \exp(\mathrm{i} \frac{\theta}{2} (\hat{X}_{N+p} \hat{Y}_{N+p\pm1} - \hat{Y}_{N+p} \hat{X}_{N+p\pm1} )),
\end{split}
\end{equation}
where $p$ and $p+N$ index the qubits representing the high- and low-spin orbitals for the spatial orbital $\phi_p$.
Here, the restriction to nearest-neighbour excitations ensures that all the Pauli-$Z$ strings cancel.
Therefore, these operators 
can be implemented with 4 CNOT gates.\cite{Anselmetti2021,Yordanov2020}
Similarly, the  Pauli-$Z$  strings also cancel for the paired two-body operators restricted to nearest-neighbour excitations,
allowing $\mathrm{exp}(\theta\,\hkap^{(2)}_{p,p\pm1})$  to be encoded with 13 CNOT gates.\cite{Anselmetti2021,Yordanov2020}
Circuit implementations for these operations are described in Ref.~\cite{Anselmetti2021}.
Furthermore, $\hkap^{(2)}_{p,p\pm1}$ could be implemented  with local qubit interactions using
a 2D qubit arrangement that provides local connectivity between the high- and low-spin states
for spatial orbitals $\phi_p$ and $\phi_{p\pm1}$, as suggested in Fig.~\ref{fig:tups}\textcolor{blue}{C}.
Consequently, each  $\hat{U}^{(m)}_{p,p\pm1}$ operator
can be implemented with 21 CNOT gates while conserving the particle number, 
$\expval*{\hat{S}^2}$ and $\expval*{\hat{S}_z}$, and anti-symmetry of $\ket{\Phi_0}$,
and the overall CNOT count for the tUPS \textit{ansatz} is
$21\, L\,(N-1)$.

\subsection{Optimal orbitals and initial qubit state}
\label{sec:oo_pp}

Orbital optimisation has been shown to
increase the accuracy of VQE for the UCC \textit{ansatz} restricted to double
excitations,\cite{Mizukami2020,Sokolov2020} pair-correlated simulations on real hardware,\cite{Zhao2023}  and separable pair approximations.\cite{Kottmann2022}
Orbital optimisation can also be used to compress the orbital space and reduce the number of qubits in a quantum simulation.\cite{Bierman2023}
Crucially, these improvements do not increase the circuit depth 
since the orbital update can be classically implemented by  transforming 
the input molecular integrals, while computing the orbital gradient requires at most the
two-body density matrix, which is already measured for energy estimation.\cite{Takeshita2020,Mizukami2020}
Therefore, we expect that orbital optimisation in the oo-tUPS \textit{ansatz} will be essential to 
maximise the accuracy of shallow quantum circuits, and we describe its
implementation in Appendix~\ref{sec:oo}.

\begin{figure}[h]
\includegraphics[width=0.9\linewidth]{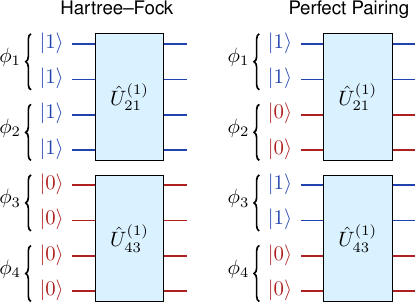}
\caption{%
\small
Different initial qubit registers with the first ``half'' layer of the tUPS 
\textit{ansatz} correspond to either the Hartree--Fock (left) or Perfect Pairing (right)
approximations.
For the Hartree--Fock case, the $\hat{U}^{(1)}_{pq}$ operators only act between 
occupied--occupied or virtual--virtual orbital pairs, and thus do not change the 
wave function.}
\label{fig:ppvshf}
\end{figure}

Since the  operators $\hat{U}^{(m)}_{p,p\pm1}$ only act between sequential 
spatial orbitals, the entanglement generated by the first tUPS layer
can be maximised by alternating the initial qubit register between occupied and 
empty spatial orbitals (Fig.~\ref{fig:ppvshf}), with no change to the circuit cost.
The first ``half'' layer is then equivalent to the  classical perfect
pairing (PP) approximation\cite{Hurley1953,Hunt1972}  as described in Appendix~\ref{sec:pp}, 
leading to the pp-tUPS \textit{ansatz}.
Since PP is a form of valence bond 
theory,\cite{Gerratt1997,Dunning2021} we expect that shallow pp-tUPS approximations 
with this alternating initial qubit register
will be most accurate using localised bonding and anti-bonding orbital pairs, 
reflecting the local qubit interactions in the circuit.
These local interactions have previously been exploited
for quantum \textit{ans\"atze} based on separable pair
approximations.\cite{Kottmann2022,Kottmann2023}
Furthermore, the relationship to PP theory suggests that the $L=1$
truncation will perform well for weakly interacting pairs of strongly correlated electrons,
while strong inter-pair correlation will require more layers.

Despite being derived from contrasting perspectives, the pp-tUPS circuit structure  (Fig.~\ref{fig:tups}\textcolor{blue}{A}) shares structural similarities to the SPA+ (separable pair approximation) approach introduced in Ref.~\cite{Kottmann2023}.
However, the pp-tUPS approach can employ multiple circuit layers, providing systematic convergence to the exact result. 
Furthermore, the circuits in Ref.~\cite{Kottmann2023} are designed using pre-defined orbital interactions based on chemical graph structures, whereas our approach is more agnostic of the molecular structure.
It is notable that the pp-tUPS approach can recover a similar set of orbitals and circuit structure to  SPA+ for linear hydrogen chains (see Section~\ref{sec:h6}), highlighting the relationship between local qubit connectivity and local orbital interactions.


\section{Results and Discussion}
\label{sec:results}

\subsection{Computational Details}
\label{sec:methods}

\begin{table*}[t!]
\caption{The number of frozen  orbitals, the Hilbert space size, and the total number of basin-hopping steps per temperature replica used 
in the BHPT optimisation for the continuous \textit{ansatz} parameters in each molecule.}
\label{tab:parameters}
\begin{ruledtabular}
\begin{tabular}{lcccc}
Molecule & Frozen core & Frozen virtuals & Hilbert Space Size & Basin-Hopping Steps per replica\\
\hline
Linear \ce{H6} & 0 & 0 & 400 & 1000 
\\
Triangular \ce{H6} & 0 & 0 & 400 & 1000 
\\
\ce{LiH} & 0 & 0 & 225 &  1000
\\
Benzene & 18 & 12 & 400 & 1000 \\
\ce{H2O} & 0 & 0 & 441 &  1000 \\
\ce{N2} & 4 & 0 & 400 &  1000 \\
\ce{BeH2} & 0 & 0 & 1225 & 250 \\
\ce{CH2} & 0 & 0 & 1225 & 1000 \\
Tetramethyleneethane &19 & 13 & 400 & 1000 \\
\end{tabular}
\end{ruledtabular}
\end{table*}

All molecular energies were computed using state-vector VQE simulations
following the protocol described in supplementary section 
``\textit{Continuous optimisation using basin-hopping}'' of Ref.~\cite{Burton2023}, 
as summarised here.
Molecular one- and two-electron integrals were computed using \textsc{PySCF}\cite{pyscf} 
for the 
molecular structures and active orbital spaces detailed in the  \textcolor{blue}{supplemental material}.\cite{supp}
Matrix representations of the Hamiltonian and the fermionic $\hkap^{(1)}_{pq}$ and $\hkap^{(2)}_{pq}$
operators were generated in the number-preserving and $\expval*{S_z}=0$ Hilbert space using
\textsc{OpenFermion}.\cite{OpenFermion}
VQE calculations 
were then simulated using a developmental version of the GMIN global optimisation program.\cite{gmin}
The continuous  parameters  of the tUPS, oo-tUPS, and pp-tUPS \textit{ans\"atze} were optimised using basin-hopping parallel tempering\cite{Li1987,Wales1997,StrodelLWW10}  (BHPT) to efficiently search for 
the global minimum, as summarised for VQE simulations in Ref.~\cite{Burton2023}.
This BHPT scheme included eight replica basin-hopping calculations with temperatures distributed exponentially between $0.0001\,\mathrm{E_h}$
and $0.01\,\mathrm{E_h}$, and exchange between replicas was attempted with a mean frequency of 10 steps.
On each basin-hopping step, the L-BFGS\cite{Nocedal1980,Broyden1970,Fletcher1970,Goldfarb1970,Shanno1970}  
optimisation procedure was used with analytic gradients, and the convergence criteria was set 
to a root-mean-square gradient value below $10^{-6}\,\mathrm{E_h}$
with a maximum of 2000 steps.
The number of frozen  orbitals, the Hilbert space size, and the total number of basin-hopping steps per replica  is tabulated  in Table \ref{tab:parameters}.

\begin{figure}[b!]
	\vspace{1em}
	\includegraphics[width=0.95\linewidth]{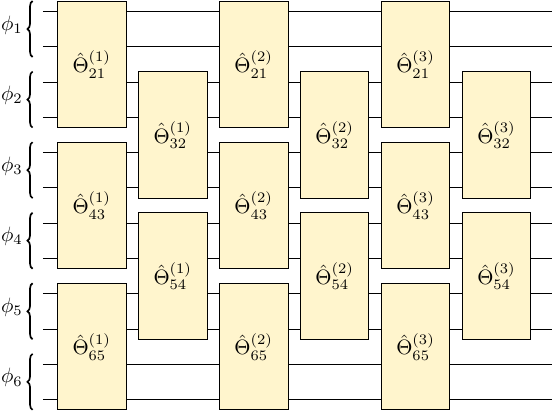}
	\caption{Structure of the orbital transformation circuit in the current VQE simulations.
		This circuit is applied after the tUPS \textit{ansatz}, with 
		$\hat{\Theta}_{p+1,p}^{(m)} = \mathrm{exp}(\theta_{p+1,p}^{(m)} \hkap_{p+1,p}^{(1)})$
		and a unique parameter $\theta_{p+1,p}^{(m)}$ for each repetition $m$.
	} 
	\label{fig:s1}
\end{figure}

While orbital optimisation can be implemented by transforming the molecular integrals, it is easier to include orbital optimisation in our current implementation using additional 
one-body operators that are optimised as part of the \textit{ansatz}.
These unitary operations are arranged in a tiled circuit structure, as illustrated in Fig.~\ref{fig:s1}, 
and applied after the primary tUPS \textit{ansatz} structure (i.e., at the end of the state preparation circuit).
The total number of one-body operators required is $\frac{1}{2}N(N-1)$, which dictates the number of layers in the orbital transformation circuit. 
This approach has the same expressibility as transforming the molecular integrals, and
a similar approach is used for  hardware experiments in Ref.~\cite{Rubin2020}.
These operators are excluded from CNOT counts because the orbital optimisation could be performed using classical pre-processing.

The alternating initial qubit state was prepared by applying a suitable series of paired two-body operators
to the HF ground state.
For triplet calculations, the initial state was prepared by applying the 
triplet excitation operator\cite{HelgakerBook}
$\hat{T}_{pq} = \frac{1}{\sqrt{2}}( \hat{p}^{\dagger} \hat{q}^{\vphantom{\dagger}}
- \hat{\bar{p}}^{\dagger} \hat{\bar{q}}^{\vphantom{\dagger}})$
to the doubly-occupied HOMO-LUMO orbital pair. 
Since triplet calculations were always performed with orbital optimisation, 
the initial orbitals were not adjusted prior to VQE optimisation.

Fermionic-excitation-based (FEB) and qubit-excitation-based (QEB) 
ADAPT-VQE simulations were performed with the \textsc{QForte} program.\cite{QForte}
The operator pool contained all generalised one- and two-body fermionic or qubit excitation 
operators without any spin adaption.
Calculations were performed up to a maximum of 200 operators and the  
number of CNOT gates was computed using the circuit implementations provided in 
Ref.~\cite{Yordanov2021}.

\subsection{Characterising the tUPS \textit{ansatz} properties}
\label{sec:h6}

We performed numerical VQE simulations to examine the accuracy of the tUPS \textit{ansatz} structure, 
as detailed in Section~\ref{sec:methods}. 
The faster convergence with respect to the number of layers of the oo-tUPS and pp-tUPS approximations
is demonstrated in
the linear and triangular isomorphs of \ce{H6} (STO-3G) with a nearest-neighbour separation of
$R(\ce{H-H})=\SI{2.0}{\angstrom}$.
For the linear structure, this bond length lies between the weakly correlated (equilibrium) and 
the strongly correlated (dissociation) regimes.
The triangular structure corresponds to a spin-frustrated lattice,
with no configuration where only opposite-spin electrons occur on neighbouring atoms.

\begin{figure*}[htb!]
\includegraphics[width=0.95\linewidth]{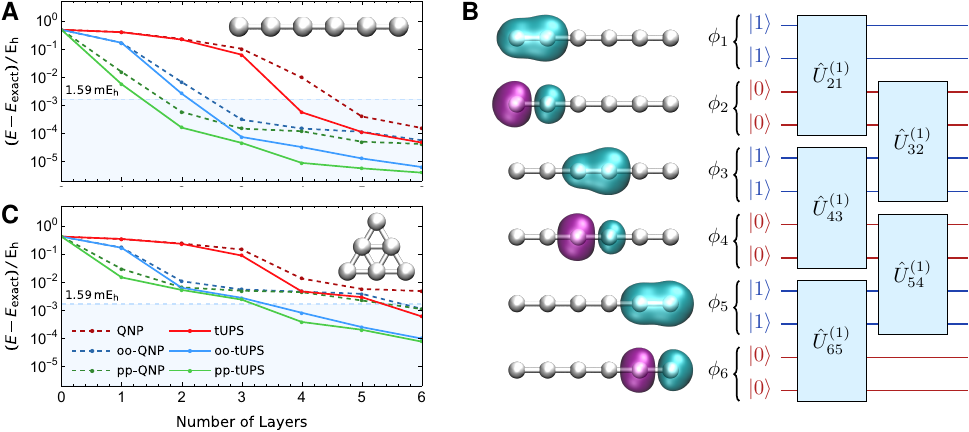}
\caption{\small
The tUPS, oo-tUPS, and pp-tUPS methods accelerate the energy convergence with 
respect to the number of layers
for (\textbf{\textsf{A}}) linear \ce{H6} and  (\textbf{\textsf{C}}) triangular \ce{H6} with $R(\ce{H-H})=\SI{2.0}{\angstrom}$.
(\textbf{\textsf{B}}) The optimal orbitals of the pp-tUPS \textit{ansatz} ($L=1$) for linear \ce{H6} 
form localised bonding and anti-bonding orbital pairs, highlighting the importance of local  
interactions.
Note that the pp-tUPS \textit{ansatz} is invariant to rotations of the orbitals within each pair.
Molecular geometries are provided in the \textcolor{blue}{supplemental material}.\cite{supp}
}
\label{fig:h6}
\end{figure*}

First, we compare the tUPS accuracy
with the original QNP approach\cite{Anselmetti2021} for a given $L$,
noting that each QNP layer contains 10 operators and 85 CNOTs compared to 15 operators and 105 CNOTs for the tUPS \textit{ansatz}.
Using the ground-state Hartree--Fock (HF) orbitals for linear \ce{H6}, the QNP approximation requires $L=5$ to
obtain chemical accuracy ($1.59\,\mathrm{mE_h}$),
while the tUPS approach requires only $L=4$ (Fig.~\ref{fig:h6}\textcolor{blue}{A}).
The greater flexibility of the $\hat{U}^{(m)}_{pq}$ operators in the tUPS \textit{ansatz} is essential when the
wave function becomes increasingly entangled, providing lower energies than QNP for deeper circuits.
Including orbital optimisation significantly improves the accuracy for 
shallow circuits, deviating from the exact result by only $10^{-4}\,\mathrm{E_h}$ for oo-tUPS with $L=3$.
Furthermore, the pp-tUPS approach requires only $L=2$ to reach chemical 
accuracy, reducing the deviation in the energy by a factor of $10^3$ compared to the HF-based tUPS 
\textit{ansatz} with the same circuit depth.
A similar reduction in the number of layers can be achieved using oo- or pp- variants of the QNP \textit{ansatz}, although the oo- and pp-tUPS approaches provide as much as a $10\times$ greater accuracy than their oo- or pp-QNP counterparts for deep circuits.
Consequently, our approach reduces the number 
of operators required to reach chemical accuracy in linear \ce{H6} 
from 50 to 30 with the QNP and pp-tUPS methods, respectively, 
and the number of CNOT gates from 425 to 210.

While the wave function will be orbital-independent in the $L\rightarrow\infty$ limit, truncated approximations 
will depend on the choice of molecular orbitals.
The optimal pp-tUPS orbitals for linear \ce{H6} with $L=1$ form pairs of localised bonding and anti-bonding orbitals between 
alternating bonds in the molecule (Fig.~\ref{fig:h6}\textcolor{blue}{B}). 
These optimal orbitals are very similar to the localised \ce{H4} orbitals used in circuits based on molecular graphs introduced in Ref.~\cite{Kottmann2023}.
Their localised structure illustrates the close relationship between quantum approximations with local 
qubit connectivity and valence bond theory, providing physical intuition into how these quantum \textit{ans\"atze} 
capture electron correlation.
This pairing-based intuition suggests that the pp-tUPS approximation can provide chemically-accurate energies with $L=2$
because the circuit structure can capture both intra- and inter-pair correlations for this system with 
three pairs of strongly interacting electrons.

The spin-frustrated triangular \ce{H6} structure exhibits stronger electron correlation,
with many near-degenerate configurations that provide significant contributions to the ground state.
The original QNP approach\cite{Anselmetti2021} fails to reach 
chemical accuracy within 6 layers, while oo- and pp- variants of the QNP \textit{ansatz} require $L=6$ (Fig.~\ref{fig:h6}\textcolor{blue}{C}).
In contrast, the standard tUPS \textit{ansatz} achieves chemical accuracy with $L=6$, demonstrating that the greater flexibility of 
$\hat{U}^{(m)}_{pq}$ [Eq.~\eqref{eq:ugate}] is vital for strongly entangled states, where 
arbitrary unitary transformations are required.
The oo-tUPS and pp-tUPS methods provide further improvement, with $L=4$ sufficient to reach chemical accuracy in 
both cases.
Although this system cannot be easily decomposed into weakly-interacting pairs of electrons, the pp-tUPS
initial qubit state still improves the accuracy of shallow circuits compared to oo-tUPS by maximising the 
correlation captured by the first layer of the \textit{ansatz}.

\subsection{Achieving gate efficiency}

The practicality of \textit{ansatz} preparation on real quantum hardware is dominated 
the number of two-qubit CNOT gates, which provide the greatest contribution to the circuit noise.
We investigated the accuracy and gate-efficiency of the tUPS approximation using the linear and triangular \ce{H6} structures, 
the weakly correlated \ce{LiH} molecule, and the delocalised $\uppi$-system in benzene.
We assume that spin-adapted one-body and paired two-body operators in the tUPS \textit{ansatz} require
4 and 13 CNOT gates, respectively, using the circuit implementations described in Ref.~\cite{Yordanov2020}.

\begin{figure*}[ht!]
\includegraphics[width=0.95\linewidth]{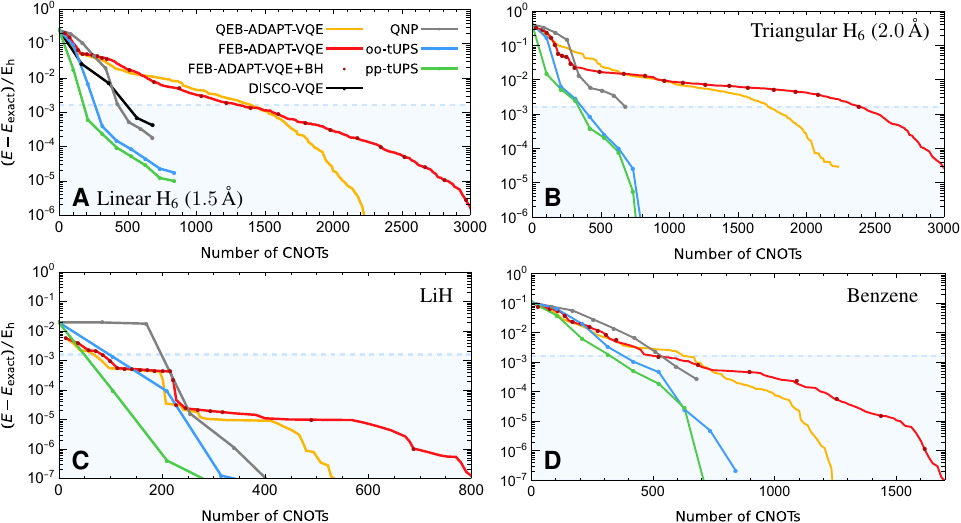}
\caption{\small 
The oo- and pp-tUPS methods significantly reduce the quantum resources required to quantitatively 
predict strongly correlated molecular energies compared to adaptive optimisation methods. 
The accuracy of the energy for a given number of CNOT gates is shown for 
(\textbf{\textsf{A}}) linear \ce{H6} with $R(\ce{H-H})=\SI{1.5}{\angstrom}$,
(\textbf{\textsf{B}}) triangular \ce{H6} with $R(\ce{H-H})=\SI{2.0}{\angstrom}$, 
(\textbf{\textsf{C}}) \ce{LiH} with $R(\ce{Li-H})=\SI{1.546}{\angstrom}$, and
(\textbf{\textsf{D}}) benzene using the (6e,~6o) active space at the experimental geometry, 
provided in the \textcolor{blue}{supplemental material}.\cite{supp}
DISCO-VQE results are taken from Ref.~\cite{Burton2023} and are only available for linear \ce{H6}.
}
\label{fig:efficiency}
\end{figure*}

To compare with previous results, we first consider  linear \ce{H6}  at $R(\ce{H-H})=\SI{1.5}{\angstrom}$.
\cite{Yordanov2021,Grimsley2019,Burton2023}
The fermionic-excitation-based\cite{Grimsley2019} (FEB) or qubit-excitation-based\cite{Yordanov2021} (QEB) variants 
of ADAPT-VQE
 require 1326 and 1410 CNOT gates to reach chemical accuracy, respectively,  (Fig.~\ref{fig:efficiency}\textcolor{blue}{A})
while similar calculations using the selected projective quantum eigensolver (SPQE) 
 converge with at least 4000 CNOT gates using the STO-6G basis set.\cite{Stair2021,Magoulas2023a,Magoulas2023b}
In contrast, only 568 CNOT gates are required if discrete optimisation is used to select
the best operator sequence from a pool containing all spin-adapted one-body and paired two-body operators.\cite{Burton2023}
Remarkably, the pp-tUPS \textit{ansatz} outperforms these adaptive methods
and provides chemical accuracy with only 210 CNOT gates, giving an 84\,\% reduction relative to FEB-ADAPT-VQE. 
Orbital optimisation and the alternating initial qubit state are essential for achieving this gate efficiency, 
as demonstrated by comparing to the oo-tUPS and QNP approaches, which require 315 and 510 CNOT gates to reach 
chemical accuracy, respectively.  
Therefore, the pp-tUPS approach sets a new standard for the number of two-qubit CNOT gates
required to obtain a chemically accurate quantum circuit for linear \ce{H6}.

Since the triangular \ce{H6} structure features stronger correlation than linear \ce{H6},
adaptive methods such as ADAPT-VQE require 
more operators and CNOT gates to reach chemical accuracy, with 2402 and 1726 CNOT gates
required for FEB-ADAPT-VQE and QEB-ADAPT-VQE, respectively (Fig.~\ref{fig:efficiency}\textcolor{blue}{B}).
In contrast, the oo-tUPS and pp-tUPS approximations provide chemical accuracy with only 420 CNOT gates, 
giving an 82.5\,\% reduction compared to FEB-ADAPT-VQE.
Therefore, the  pp-tUPS \textit{ansatz} can describe weak and strong correlation with a similar quantum 
resource cost.

The \ce{LiH} and benzene molecules have been extensively studied using
classical and quantum algorithms.
At equilibrium, \ce{LiH} is dominated by a single Slater determinant and 
both oo-tUPS and pp-tUPS provide chemical accuracy with $L=1$, corresponding to 105 CNOT gates 
(Fig.~\ref{fig:efficiency}\textcolor{blue}{C}).
Since \ce{LiH} does not feature particularly strong correlation, the FEB- and QEB-ADAPT-VQE approaches provide similar 
gate efficiency to the pp-tUPS, although the pp-tUPS approximation converges more rapidly once the energy is within $10^{-4}\,\mathrm{E_h}$ 
of the  exact result.
The pp-tUPS approach reaches chemical accuracy for benzene with 420 CNOT gates, compared to 524 for FEB-ADAPT-VQE, 
providing a reduction of 20\,\% (Fig.~\ref{fig:efficiency}\textcolor{blue}{D}).
Consequently, the pp-tUPS approach significantly reduces the number of two-qubit CNOT gates required for chemically
accurate predictions of both weakly and strongly correlated molecular energies, while also 
preserving the particle number, Pauli antisymmetry, $\expval*{\hat{S}_z}$, and $\expval*{\hat{S}^2}$ symmetries  
of the initial state, and avoiding adaptive optimisation.

To ensure a fair comparison between ADAPT-VQE and tUPS, the basin-hopping procedure was also used to
optimise the \textit{ansatz} discovered by  ADAPT-VQE after a certain number of steps (dark red dots in Fig.~\ref{fig:efficiency}).
In each case, there is negligible improvement in the energy optimised using this ADAPT-VQE-BH approach compared to the standard ADAPT-VQE result,
demonstrating that the improved accuracy of the oo- and pp-tUPS hierarchy arises from the \textit{ansatz} structure rather the choice of optimisation algorithm for the continuous parameters. 
These results are consistent with Ref.~\cite{Grimsley2023}, which suggests that ADAPT-VQE typically finds a solution close to the global minimum if the continuous parameters are recycled on each macro-iteration.

\subsection{Predicting accurate potential energy surfaces}

\begin{figure*}[htb]
	\includegraphics[width=0.95\linewidth]{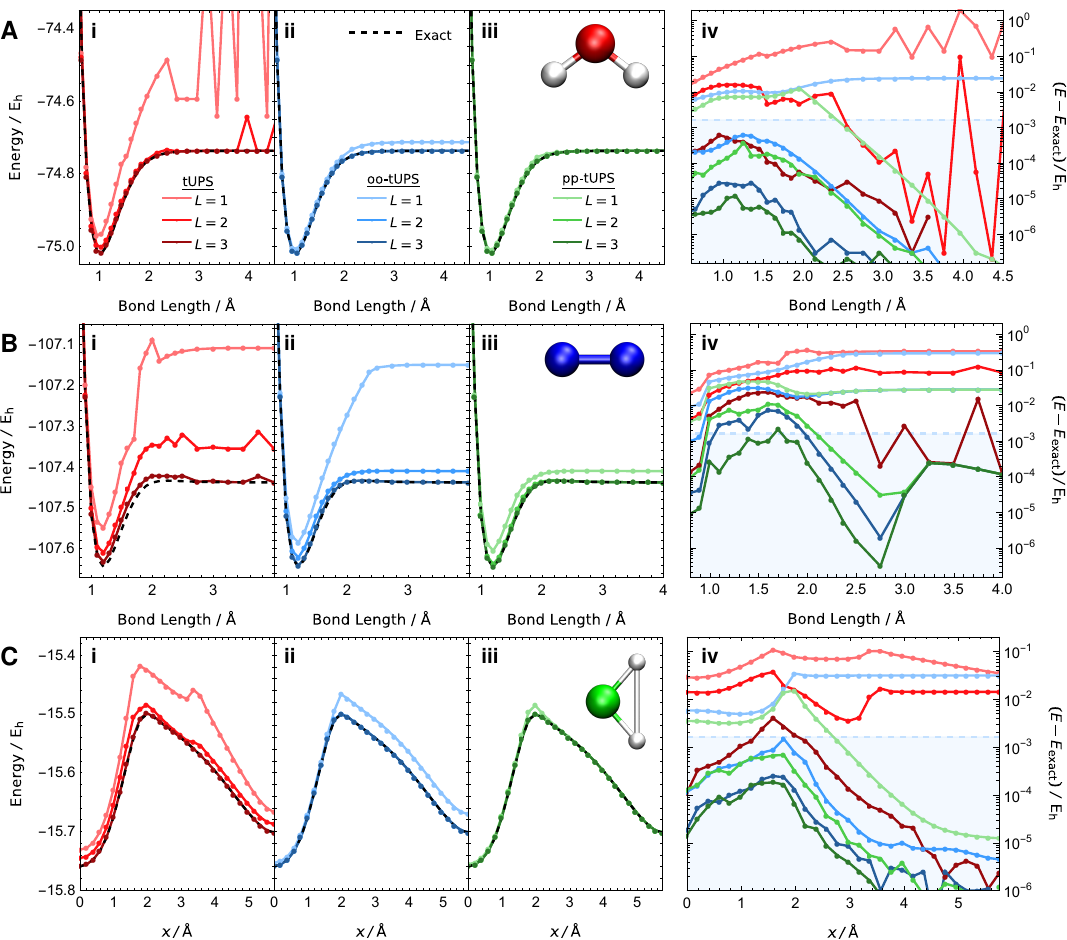}
	\caption{\small Orbital optimisation is essential for computing smooth potential energy surfaces. 
    Energy surfaces are computed for the \ce{H2O} (STO-3G) symmetric stretch (\textbf{\textsf{A}}), \ce{N2} (STO-3G; 4 frozen core orbitals) dissociation (\textbf{\textsf{B}}),
    and  \ce{Be + H2} (STO-6G) insertion pathway (\textbf{\textsf{C}}) using the tUPS (\textbf{\textsf{i}}), oo-tUPS (\textbf{\textsf{ii}}), and pp-tUPS (\textbf{\textsf{iii}}) 
    approximations, with the energetic accuracy compared in (\textbf{\textsf{iv}}).
    Molecular geometries are detailed in the \textcolor{blue}{supplemental material}.\cite{supp}}
	\label{fig:pes}
\end{figure*}


Chemical simulations rely on accurate predictions of molecular potential energy surfaces.
However, classical methods struggle to balance the different correlation 
that occurs as molecular structures change, such as competing electronic configurations 
and spin-coupling during chemical reactions.
We assessed the accuracy of tUPS, oo-tUPS, and pp-tUPS approximations for archetypal potential energy surfaces, 
including the  dissociation of \ce{H2O} and \ce{N2},
and the insertion of \ce{Be} into \ce{H2}. 

Smooth potential energy surfaces are required to compute 
nuclear forces for geometry optimisation or dynamic simulations.
Orbital optimisation in the oo-tUPS or pp-tUPS methods 
is essential to obtain smooth energy surfaces for all the molecules considered (Fig.~\ref{fig:pes}).
In contrast, the HF-based tUPS \textit{ansatz} ``jumps'' between different solutions, 
despite using basin-hopping optimisation to identify the global minimum with respect to the continuous parameters.
These jumps are worst for strongly correlated states, such as the molecular dissociation limit.

The accuracy of the oo-tUPS and pp-tUPS \textit{ansatz} for dissociated \ce{H2O} and \ce{N2}
provides intuition into how electron correlation is captured.
The symmetric stretch of \ce{H2O} simultaneously breaks two bonds into two spatially separated pairs of 
spin-coupled electrons.
The pp-tUPS approximation accurately captures this correlation with $L=1$ because the first half-layer strongly
couples the electrons within each pair, and the second half-layer introduces the inter-pair coupling 
(Fig.~\ref{fig:pes}\textcolor{blue}{A}). 
In contrast, the oo-tUPS \textit{ansatz} requires $L=2$ to predict the correct dissociation limit since the HF-style 
initial qubit register is less efficient at describing the spin-coupled electron pairs.
Similarly, breaking the \ce{N2} triple bond gives three pairs of spin-coupled electrons and
the pp-tUPS approximation requires $L=2$ to capture the inter-pair correlation, while the oo-tUPS approach 
accurately predicts the dissociation limit with $L=3$ (Fig.~\ref{fig:pes}\textcolor{blue}{B}). 
These results suggest that the pp-tUPS method can dissociate $L$ bonds using $L-1$ layers,
providing valuable intuition into the chemical applicability of this quantum \textit{ansatz}.

Reaching chemical accuracy across a full potential energy surface with a consistent and small number of unitary 
operators is particularly challenging for adaptive optimisation methods, which typically require more operators
for intermediate bond lengths or the strongly correlated dissociation limit.\cite{Yordanov2021,Grimsley2019,Magoulas2023a,Magoulas2023b}
In contrast, discrete global optimisation of the operator sequence showed that accurate binding 
curves for \ce{H2O} and \ce{N2} can be achieved with a constant number of operators.\cite{Burton2023}
Using the pp-tUPS \textit{ansatz}, two layers (36 operators; 252 CNOTs) are sufficient to get a chemically accurate binding curve for 
\ce{H2O}, while three layers (45 operators; 315 CNOTs) are required for \ce{N2}.
By comparison, analogous FEB-ADAPT-VQE calculations required around 
400 and 2000 CNOT gates to reach chemical accuracy for equilibrium and stretched \ce{N2}, respectively.\cite{Tsuchimochi2022}
The ability of pp-tUPS to give chemical accuracy for different geometries with a consistent number of CNOT 
gates will be essential for balancing the quantum hardware noise along potential energy surfaces.

Compared to the spin-coupling correlation for bond dissociation,
the \ce{Be +H2} insertion mechanism features strong mixing between two dominant closed-shell configurations.\cite{Purvis1982}
We consider the reaction trajectory defined in Ref.~\cite{Evangelista2011} with the STO-6G basis used in 
Ref.~\cite{Magoulas2023a}.
Like the \ce{H2O} and \ce{N2} binding curves, orbital optimisation is essential to obtain a smooth potential energy surface (Fig.~\ref{fig:pes}\textcolor{blue}{C}).
However, one layer of the oo-tUPS \textit{ansatz} is not sufficient to get a balanced binding curve, 
giving less accurate energies for the dissociated regime where there is competition between 
the \ce{Be} $\mathrm{(1s)^2(2s)^2}$ and $\mathrm{(1s)^2(2p)^2}$ configurations\cite{Mok1996} (Fig.~\ref{fig:pes}\textcolor{blue}{C}).
The pp-tUPS \textit{ansatz} provides qualitative accuracy with $L=1$ (18 operators; 126 CNOTs) and 
reaches chemical accuracy at all points with $L=2$ (36 operators; 252 CNOTs).
The highly-accurate dissociation limit of pp-tUPS  with $L=1$ can be rationalised as there are two correlated pairs of electrons
on the \ce{Be} atom, and the overall system is a direct product of the \ce{Be} and \ce{H2} wave functions.
Crucially, the pp-tUPS \textit{ansatz} achieves this accuracy with a consistent wave function structure, while 
adaptive techniques typically select between $\times$5 to $\times$130 more operators  at the 
crossing point ($x\sim\SI{2}{\angstrom}$)  compared to the dissociation limit.\cite{Magoulas2023a}

\subsection{Computing spin-state energetics}
Resolving the energies of different spin states, such as as singlet-triplet gaps, is important
for developing efficient organic light-emitting diodes, singlet fission, and photocatalysis.
However, current quantum algorithms can only compute spin energetics using excited-state methods
such as variational quantum deflation,\cite{Ibe2022} 
constraining $\expval*{\hat{S}^2}$ using Lagrange multipliers,\cite{Shirai2023} 
or subspace expansions such as the nonorthogonal VQE\cite{Baek2023} 
and quantum equation-of-motion methods.\cite{Asthana2023,Colless2018,Ollitrault2020,Matousek2023,Gao2021}
Since the tUPS \textit{ansatz} only contains operators that commute with $\hat{S}_z$ and $\hat{S}^2$,
the spin of the initial state is conserved for all truncations.
Therefore, different spin-state energies can now be computed on an equal footing using suitable
initial states, without any modifications to the VQE optimisation.

\begin{figure}[t]
	\includegraphics[width=\linewidth]{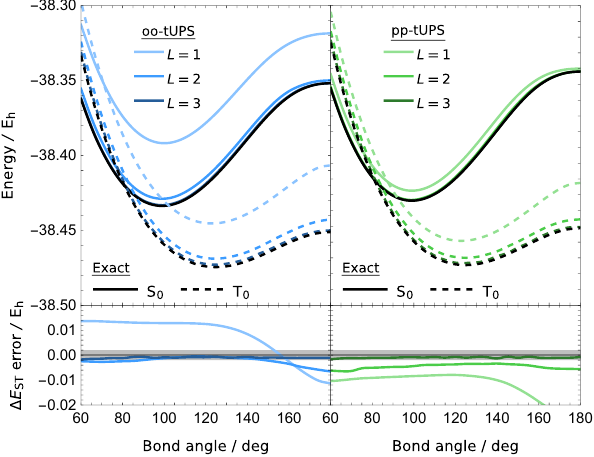}
	\caption{\small The oo-tUPS and pp-tUPS \textit{ans\"atze} accurately predict the singlet-triplet gap 
		of methylene (STO-3G) with $R(\ce{C-H}) = \SI{1.117}{\angstrom}$ by preserving the  $\expval*{\hat{S}_z}$ and $\expval*{\hat{S}^2}$ quantum numbers of the initial state.}
	\label{fig:ch2}
\end{figure}

\begin{figure}[b!]
	\includegraphics[width=\linewidth]{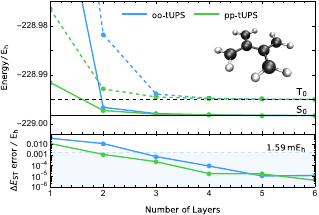}
	\caption{\small The oo-tUPS and pp-tUPS \textit{ans\"atze} systematically converge to the 
		exact $\mathrm{S_0}$ and $\mathrm{T_0}$ energies for planar TME as more layers are added to the \textit{ansatz}. 
		The $\uppi$-system (6e, 6o) active space is used with the STO-3G basis set at the molecular geometry provided in the  \textcolor{blue}{supplemental material}.\cite{supp}
		The singlet-triplet gap ($\Delta E_{\text{ST}}$) can be predicted within chemical accuracy using two layers (30 operators; 210 CNOTs) of the pp-tUPS \textit{ansatz}. }
	\label{fig:tme}
\end{figure}

\begin{figure*}[htb]
\includegraphics[width=0.95\linewidth]{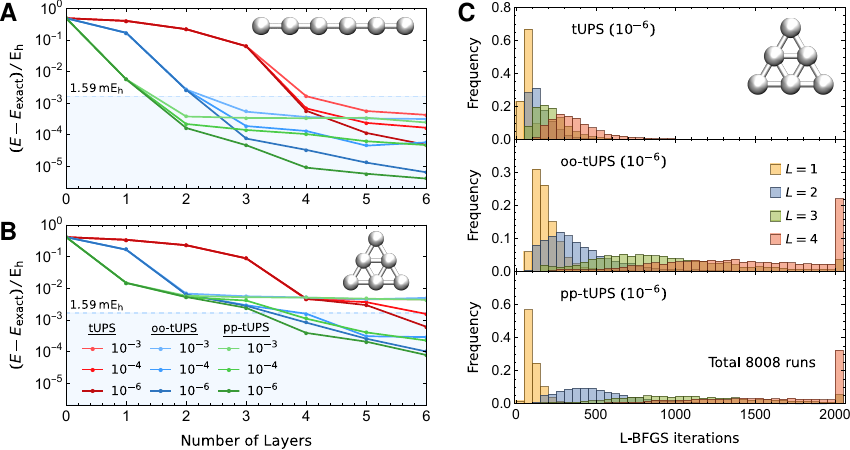}
\caption{%
\small
Comparison of the optimised ground-state energies obtained for the tUPS hierarchy with different gradient RMS convergence thresholds for (\textbf{\textsf{A}})  linear \ce{H6} and  (\textbf{\textsf{B}}) triangular  \ce{H6} with $R(\ce{H-H}) = \SI{1.5}{\angstrom}$.  (\textbf{\textsf{C}})
The number of steps required to reach convergence for individual L-BFGS runs in triangular \ce{H6} as more circuit layers are added or orbital optimisation is included.}
\label{fig:opt}
\end{figure*}

We first illustrate this approach using the bending mode of methylene, 
which involves a crossing of the $\mathrm{S_0}$ and $\mathrm{T_0}$ states.
Both the oo-tUPS and pp-tUPS approximations qualitatively reproduce  the singlet-triplet intersection point
with $L=2$ (36 operators; 252 CNOTs) and can predict the singlet-triplet gap 
$\Delta E_\text{ST}=E_\text{S} - E_\text{T}$ 
to within chemical accuracy at all points with $L=3$ (54 operators; 378 CNOTs), as shown 
in Fig.~\ref{fig:ch2}.
Like the ground-state potential energy surfaces, the pp-tUPS is more accurate than oo-tUPS  for $L=1$.
However, this improvement is less significant for the $\mathrm{T_0}$ state since the pp-tUPS \textit{ansatz} cannot 
capture any additional correlation between the triplet-coupled electrons.

Tetramethyleneethane (TME) is a more challenging disjoint diradical, where the degenerate molecular
orbitals are spatially separated.\cite{Pozun2013,Veis2018}
The (6e, 6o) active space corresponding to the carbon $\pi$-system provides the natural approximation for predicting
$\Delta E_{\text{ST}}$. 
Both the oo-tUPS and pp-tUPS approaches systematically converge to the exact $\mathrm{S_0}$ and $\mathrm{T_0}$
energies as the number of \textit{ansatz} layers is increased (Fig.~\ref{fig:tme}).
The pp-tUPS approach provides a balanced representation of the two states and can predict the singlet-triplet energy
gap within chemical accuracy using two-layers (30 operators; 210 CNOTs).
Again, the oo-tUPS approximation is less accurate for shallow circuits, but quantitatively predicts the singlet-triplet
gap with $L=3$.
Consequently, the tUPS \textit{ansatz} allows different spin states to be directly targeted in the 
VQE formalism, which was previously challenging using non-symmetry-preserving adaptive optimisation techniques.

\subsection{Analysis of Numerical Optimisation}

The number of quantum measurements to evaluate the VQE gradient to a precision of $\epsilon$ scales as $\mathcal{O}(\epsilon^{-2})$.\cite{Tilley2022}
Therefore, it is preferable to make the RMS gradient convergence threshold as loose as possible while still retaining chemical accuracy in the final energy.
In Fig.~\ref{fig:opt}\textcolor{blue}{A} and \ref{fig:opt}\textcolor{blue}{B}, we compare the converged
tUPS, oo-tUPS, and pp-tUPS energies for the linear and triangular \ce{H6} systems using different RMS gradient convergence thresholds of $10^{-3}$, $10^{-4}$ and $10^{-6}\,\mathrm{E_h}$.
The number of layers required to reach chemical accuracy is unchanged in the linear \ce{H6} system, 
although the overall accuracy obtained with deep quantum circuits is reduced for the looser convergence 
threshold. 
In the \ce{H6} triangle, a convergence  threshold of  $10^{-3}\,\mathrm{E_h}$ prevents chemical accuracy from 
being obtained, but a threshold of $10^{-4}\,\mathrm{E_h}$ is sufficient to leave the conclusions unchanged.

It is well known that variational quantum algorithms are prone to challenging numerical optimisation.
The basin-hopping algorithm performs many L-BFGS runs from randomly perturbed
starting points, providing data to assess the convergence properties of the tUPS \textit{ansatz} hierarchy.
For the \ce{H6} triangle, we see that the number of L-BFGS steps required for convergence generally increases 
with more circuit layers for all variants of the tUPS hierarchy (Fig.~\ref{fig:opt}\textcolor{blue}{C}).
The convergence becomes slower when orbital optimisation is included in the 
oo- and pp-tUPS \textit{ans\"atze}.
For $L=4$, 20-30\,\% of L-BFGS runs reach the maximum number of 2000 steps, giving states that are variationally lower in energy but are not fully converged.
This large number of L-BFGS steps is concordant with QNP calculations in Ref.~\cite{Anselmetti2021}, and is illustrative 
of the optimisation challenges in VQE calculations.

\section{Conclusions}
We have shown that physically-accurate parametrisations for electronic states can be constructed  
with very shallow quantum circuits using the fixed pp-tUPS \textit{ansatz}, avoiding expensive adaptive 
optimisation methods.
This \textit{ansatz} is systematically improvable, and  combines physical accuracy 
with gate efficiency while conserving particle number, fermionic antisymmetry, and the $\expval*{\hat{S}_z}$
and $\expval*{\hat{S}^2}$ quantum numbers of the initial state.
The accuracy that can be achieved with shallow quantum circuits is maximised by incorporating orbital optimisation 
and an initial qubit register that is derived from perfect pairing valence bond theory.
Numerical simulations on molecules with strongly correlated electronic states demonstrate that chemically
accurate potential energy surfaces and singlet-triplet gaps can be predicted with significantly fewer two-qubit CNOT gates compared to state-of-the-art adaptive optimisation methods.

Previously, adaptive optimisation has been essential to identify a physically accurate 
sequence of unitary operators with shallow quantum circuit implementations.\cite{Grimsley2019,Burton2023,Yordanov2021}
In contrast, the pp-tUPS \textit{ansatz} achieves accuracy and gate efficiency by using fermionic 
operators that only act between nearest-neighbour spatial orbitals and by considering their 
properties as Lie algebraic generators for unitary transformations.
This shift away from adaptive \textit{ansatz} design offers key advantages 
for practical quantum simulations: 
it avoids the measurement costs required to adaptively optimise the operator sequence;
and gives a consistent circuit structure with a well-defined quantum resource cost across all molecular structures.
These advances lay the foundation for new high-accuracy fixed approximations for 
electronic states that embrace the natural functionality of qubit rotations, without sacrificing fundamental 
physical symmetries.

This work has considered the full expressibility of the tUPS \textit{ansatz} hierarchy using 
global optimisation techniques.
Our current approach has focussed on finding the global minimum, but requires a large number of energy and gradient evaluations.
Future investigations will be required to understand the efficiency of numerically optimising the energy, 
including the best optimisation algorithm, the dependence on the initial guess, and the structure of the underlying energy landscape.
Suitable initial guesses may be identified using concepts from valence bond theory, potentially by taking advantage of related quantum \textit{ans\"atze} such as Refs.~\cite{Kottmann2022,Kottmann2023,Kottmann2024,Ghasempouri2023} or using established orbital localisation techniques.\cite{Pipek1989,Foster1960}
However, it is not clear whether the \textit{ansatz} structure, or the inclusion of orbital optimisation,
will improve or worsen the issues of local minima and barren plateaus in variational 
quantum algorithms.\cite{Bittel2021,McClean2018}
Further numerical simulations will be necessary to assess how much the reduction in CNOT gates improves
the noise resilience of the corresponding quantum circuits,  
whether the accuracy remains consistent for larger systems, and how many circuit 
layers are needed. 

Preparing physically accurate and gate-efficient quantum representations of 
electronic states will be vital to capitalise on the functionality of near-term quantum computing.
Current state-of-the-art methods adaptively design the \textit{ansatz} for each molecule, 
but these have high quantum measurement costs, 
give different circuits along potential energy surfaces, and struggle to preserve spin symmetry.
We have presented the pp-tUPS \textit{ansatz}, which achieves both gate-efficiency
and symmetry preservation using a fixed circuit structure.
Our approach sets a new standard for the accuracy and gate efficiency that can be achieved
for strongly correlated molecules,
using as much as 84\,\% fewer two-qubit gates compared to adaptive techniques.
We believe that the tUPS \textit{ansatz} hierarchy will support the development of 
practical  simulations for strongly correlated chemistry on real quantum hardware.

\section*{Data Availability}
The numerical data required to reproduce the figures presented in this manuscript will be made available in an open-source repository upon acceptance.

\section*{Acknowledgements}
The author acknowledges insightful discussions with
Daniel Marti-Dafcik, David Tew, Paul Johnson, Alex Thom, and Pierre-Fran\c{c}ois Loos. 
The author also thanks Erik Kjellgren for highlighting  improvements to the parameter-shift rules proposed in the original draft.
HGAB was supported by Downing College, Cambridge through the Kim and Julianna Silverman Research Fellowship.

\appendix

\section{Relating fermionic operators to Given's rotations}
\label{sec:givens}
The action of $\mathrm{exp}(\theta \hkap_{pq}^{(1)})$ and  $\mathrm{exp}(\theta \hkap_{pq}^{(2)})$ for many-electron systems
can be understood by considering how these operators transform many-body Slater determinants. 
For a given $(p,q)$ pair, we can group determinants according to their occupation 
of the orbitals  $\phi_p$ and $\phi_q$.
In what follows, an arbitrary determinant is denoted $\ket{k_q k_p  k_{\bar{q}}  k_{\bar{p}} ; \bm{k}'}$, 
where $k_p$ ($k_{\bar{p}}$) are the occupation numbers of the high (low) spin orbital corresponding to $\phi_p$,
and $\bm{k}'$ is the occupation vector for orbitals excluding $p$ and $q$.
Since neither  $\hkap_{pq}^{(1)}$ or $ \hkap_{pq}^{(2)}$ can change the occupation of any orbital except $\phi_p$ or $\phi_q$,
matrix representations for each $\bm{k}'$  can be expressed as
\setcounter{MaxMatrixCols}{20}
\begin{widetext}
\begin{equation}
\hkap_{pq}^{(1)}
=
\begin{pmatrix}
 0 &    &    &    &    &    &    &    &    &    &    &    &    &    &    &    \\
   &  0 & -1 &    &    &    &    &    &    &    &    &    &    &    &    &    \\
   &  1 &  0 &    &    &    &    &    &    &    &    &    &    &    &    &    \\
   &    &    &  0 & -1 &    &    &    &    &    &    &    &    &    &    &    \\
   &    &    &  1 &  0 &    &    &    &    &    &    &    &    &    &    &    \\
   &    &    &    &    &  0 &    &    &    &    &    &    &    &    &    &    \\
   &    &    &    &    &    &  0 & -1 & -1 &  0 &    &    &    &    &    &    \\
   &    &    &    &    &    &  1 &  0 &  0 & -1 &    &    &    &    &    &    \\
   &    &    &    &    &    &  1 &  0 &  0 & -1 &    &    &    &    &    &    \\
   &    &    &    &    &    &  0 &  1 &  1 &  0 &    &    &    &    &    &    \\
   &    &    &    &    &    &    &    &    &    &  0 &    &    &    &    &    \\
   &    &    &    &    &    &    &    &    &    &    &  0 & -1 &    &    &    \\
   &    &    &    &    &    &    &    &    &    &    &  1 &  0 &    &    &    \\
   &    &    &    &    &    &    &    &    &    &    &    &    &  0 & -1 &    \\
   &    &    &    &    &    &    &    &    &    &    &    &    &  1 &  0 &    \\
   &    &    &    &    &    &    &    &    &    &    &    &    &    &    &  0
\end{pmatrix},
\quad
\hkap_{pq}^{(2)}
=
\begin{pmatrix}
 0 &    &    &    &    &    &    &    &    &    &    &    &    &    &    &    \\
   &  0 &    &    &    &    &    &    &    &    &    &    &    &    &    &    \\
   &    &  0 &    &    &    &    &    &    &    &    &    &    &    &    &    \\
   &    &    &  0 &    &    &    &    &    &    &    &    &    &    &    &    \\
   &    &    &    &  0 &    &    &    &    &    &    &    &    &    &    &    \\
   &    &    &    &    &  0 &    &    &    &    &    &    &    &    &    &    \\
   &    &    &    &    &    &  0 &  0 &  0 & -2 &    &    &    &    &    &    \\
   &    &    &    &    &    &  0 &  0 &  0 &  0 &    &    &    &    &    &    \\
   &    &    &    &    &    &  0 &  0 &  0 &  0 &    &    &    &    &    &    \\
   &    &    &    &    &    &  2 &  0 &  0 &  0 &    &    &    &    &    &    \\
   &    &    &    &    &    &    &    &    &    &  0 &    &    &    &    &    \\
   &    &    &    &    &    &    &    &    &    &    &  0 &    &    &    &    \\
   &    &    &    &    &    &    &    &    &    &    &    &  0 &    &    &    \\
   &    &    &    &    &    &    &    &    &    &    &    &    &  0 &    &    \\
   &    &    &    &    &    &    &    &    &    &    &    &    &    &  0 &    \\
   &    &    &    &    &    &    &    &    &    &    &    &    &    &    &  0
\end{pmatrix}
\label{eq:matrixrep}
\end{equation}
where the occupation numbers $\ket{k_q k_p k_{\bar{q}} k_{\bar{p}}}$ are ordered as
\begin{equation}
\small
\{ 
\ket{0000},
\ket{1000},
\ket{0100},
\ket{0010},
\ket{0001},
\ket{1100},
\ket{1010},
\ket{1001},
\ket{0110},
\ket{0101},
\ket{0011},
\ket{1110},
\ket{1101},
\ket{1011},
\ket{0111},
\ket{1111}
\}
\end{equation}
\end{widetext}
and the remaining zeros have been omitted to highlight the block diagonal structure.
These matrix representations reveal the relationship of $\mathrm{exp}(\theta\, \hkap_{pq}^{(1)})$ and $\mathrm{exp}(\theta\, \hkap_{pq}^{(2)})$ 
to Given's rotations.\cite{Arrazola2022}
In particular, since the high- and low-spin one-body operators commute 
(i.e.\ $[\hat{\tau}_{pq} , \hat{\tau}_{\bar{p}\bar{q}} ] = 0$ where $\hat{\tau}_{pq} = \hat{p}^{\dagger} \hat{q}^{\vphantom{\dagger}} -  \hat{q}^{\dagger} \hat{p}^{\vphantom{\dagger}}$),
the one-body exponential operator corresponds to the product of two Given's rotations  as  
$\mathrm{exp}(\theta\, \hkap_{pq}^{(1)}) = \mathrm{exp}(\theta\,  \hat{\tau}_{pq} )  \mathrm{exp}(\theta\, \hat{\tau}_{\bar{p}\bar{q}} )$.
The Given's rotation $ \mathrm{exp}(\theta \, \hat{\tau}_{pq} ) $ only affects the occupancies of the high-spin orbitals corresponding to $\phi_p$ and $\phi_q$, and similarly for the
low-spin counterpart $ \mathrm{exp}(\theta \, \hat{\tau}_{\bar{p}\bar{q}} ) $.
Analogously, the paired two-body exponential operator $\mathrm{exp}(\theta \,\hkap_{pq}^{(2)})$  corresponds to a Given's rotation between 
the determinants with occupancies $\ket{k_q k_p k_{\bar{q}} k_{\bar{p}}}$ corresponding to   $\ket{1010}$ and $\ket{0101}$.\cite{Arrazola2022}

\section{Parameter shift rule for tUPS gradients}
\label{sec:shift}

Computing partial derivatives of  the energy with respect to the tUPS wave function parameters is essential for practical 
VQE simulations and gradients.
Since the energy is periodic with respect to the \textit{ansatz} parameters, a parameter-shift rule can be used to compute
analytic gradients using the same variational circuit architecture as the energy computation.\cite{Mitarai2018}
The required parameter-shift rule can be deduced by examining the eigenvalue structure of $\hkap_{pq}^{(1)}$ and 
$\hkap_{pq}^{(2)}$.
First, we consider partial derivatives with respect to a parameter defining an exponential spin-adapted one-body operator $\mathrm{exp}(\theta_1 \hkap_{pq}^{(1)})$.
From Eq.~\eqref{eq:matrixrep}, the unique eigenvalues of $\hkap_{pq}^{(1)}$ are $0$, $\pm 1\mathrm{i}$, and $\pm 2 \mathrm{i}$.
Therefore,  the exact partial derivative can be computed using an eight-point parameter shift rule.\cite{Wierichs2022} 
A convenient expression is given as
\begin{equation}
\begin{split}
\grad_{\theta_1} E=\ &\frac{1}{2}\qty[ E \qty(\theta_1 + \frac{\pi}{2}) - E\qty(\theta_1 - \frac{\pi}{2}) ] 
\\
+ \qty(\frac{2\sqrt{2}+3}{2}) &\qty[ E \qty(\theta_1 + \frac{\pi}{4}) - E\qty(\theta_1 - \frac{\pi}{4})  ]
\\
+ \qty(\frac{2\sqrt{2}-3}{2}) &\qty[ E \qty(\theta_1 + \frac{3\pi}{4}) - E\qty(\theta_1 - \frac{3\pi}{4})  ]
\\
- \frac{4\sqrt{3}}{3} &\qty[ E \qty(\theta_1 + \frac{\pi}{3}) - E\qty(\theta_1 - \frac{\pi}{3})  ]
.
\end{split}
\end{equation}
Similarly, the eigenvalues of  $\hkap_{pq}^{(2)}$ are $0$, and $\pm 2 \mathrm{i}$, and thus the exact partial derivative 
for an exponential paired two-body operator $\mathrm{exp}(\theta_2 \hkap_{pq}^{(1)})$ can be computed with 
the four-point parameter shift rule
\begin{equation}
\begin{split}
\grad_{\theta_2}E= 2&\qty[ E \qty(\theta_2 + \frac{\pi}{8}) - E\qty(\theta_2 - \frac{\pi}{8}) ] 
\\
+ \qty(1-\sqrt{2}) &\qty[ E \qty(\theta_2 + \frac{\pi}{4}) - E\qty(\theta_2 - \frac{\pi}{4})  ].
\end{split}
\end{equation}

\vspace{-0.1em}

\section{Universality of the tUPS wave function}
\label{sec:universal}

Previously, the universality of UPS wave functions constructed from spin-adapted one-body and paired two-body 
operators was derived by showing that all many-body fermionic operators can be represented as nested commutator 
expansions of $\hkap_{pq}^{(1)}$ and $\hkap_{pq}^{(2)}$.\cite{Evangelista2019,Burton2023}
The universality of the tUPS circuit structure can be shown by representing  $\hkap_{pq}^{(1)}$ and $\hkap_{pq}^{(2)}$
as nested commutators containing only operators that act between sequential spatial orbitals, i.e.\ $\hkap_{p,p\pm1}^{(1)}$ and $\hkap_{p,p\pm1}^{(2)}$.
These operators are shown to be sufficient by representing the non-sequential spin-adapted one-body and paired two-body 
operators as
\begin{subequations}
\begin{align}
\hkap_{p,p+2}^{(1)}  &= [ \hkap_{p,p+1}^{(1)},  \hkap_{p+1,p+2}^{(1)}],
\label{eq:comm1}
\\
\hkap_{p,p+2}^{(2)}  &= \frac{1}{2}  \qty[ \hkap_{p,p+1}^{(1)},  [\hkap_{p,p+1}^{(1)}, \hkap_{p+1,p+2}^{(2)}]] +  \hkap_{p+1,p+2}^{(2)}.
\label{eq:comm2}
\end{align}
\end{subequations}
Through these expressions, the exponential operators can be implemented using the identities 
$\mathrm{e}^{\hat{A} + \hat{B}} = \lim_{n\rightarrow \infty}(\mathrm{e}^{\hat{A}} \mathrm{e}^{\hat{B}})^{n}$ 
and 
$\mathrm{e}^{[\hat{A} , \hat{B}]} = \lim_{n\rightarrow \infty}(\mathrm{e}^{\hat{A}/\sqrt{n}} \mathrm{e}^{\hat{B}/\sqrt{n}} \mathrm{e}^{-\hat{A}/\sqrt{n}} \mathrm{e}^{-\hat{B}/\sqrt{n}})^{n}$.

Eq.~\eqref{eq:comm1} can be derived using the standard commutator rule $[\hE_{pq},\hE_{rs}] = \delta_{qr}\hE_{ps} - \delta_{ps}\hE_{rq}$ as
\begin{widetext}
\begin{equation}
\begin{split}
[ \hkap_{p,p+1}^{(1)},  \hkap_{p+1,p+2}^{(1)}] &= [\hE_{p,p+1} - \hE_{p+1,p}, \hE_{p+1,p+2} - \hE_{p+2,p+1} ] 
\\
&=  [\hE_{p,p+1} , \hE_{p+1,p+2}] -[\hE_{p+1,p}, \hE_{p+1,p+2}  ] -[\hE_{p,p+1} , \hE_{p+2,p+1} ] + [\hE_{p+1,p}, \hE_{p+2,p+1} ] 
\\
&=  \hE_{p,p+2}  -\hE_{p+2,p}  = \hkap^{(1)}_{p,p+2}.
\end{split}
\end{equation}
Similarly, noting that $\hkap^{(2)}_{pq} = \hat{e}_{pqpq} -  \hat{e}_{qpqp}$, where $\hat{e}_{pqrs} = \hE_{pq}\hE_{rs} - \delta_{qr}\hE_{ps}$, we can 
use the commutator identity\cite{HelgakerBook} 
$[\hE_{mn},\hat{e}_{pqrs}] = \delta_{pn}\hat{e}_{mqrs} + \delta_{rn}\hat{e}_{pqms} -   \delta_{mq}\hat{e}_{pnrs} - \delta_{ms}\hat{e}_{pqrn}$ 
to expand the nested commutator in Eq.~\eqref{eq:comm2} as
\begin{equation}
\begin{split}
\qty[ \hkap_{p,p+1}^{(1)},  [\hkap_{p,p+1}^{(1)}, \hkap_{p+1,p+2}^{(2)}]] &=
\qty[\hE_{p,p+1} - \hE_{p+1,p},  [\hE_{p,p+1} - \hE_{p+1,p},\hat{e}_{p+1,p+2,p+1,p+2} - \hat{e}_{p+2,p+1,p+2,p+1}]]
\\ &= [\hE_{p,p+1} - \hE_{p+1,p}, \hat{e}_{p,p+2,p+1,p+2} +  \hat{e}_{p+1,p+2,p,p+2} - \hat{e}_{p+2,p,p+2,p+1} + \hat{e}_{p+2,p+1,p+2,p}]
\\
&= 2 (  \hat{e}_{p,p+2,p,p+2} - \hat{e}_{p+2,p,p+2,p}  ) - 2 (  \hat{e}_{p+1,p+2,p+1,p+2} - \hat{e}_{p+2,p+1,p+2,p+1}  ) 
\\
&= 2 \hkap_{p,p+2}^{(2)} - 2 \hkap_{p+1,p+2}^{(2)} 
\end{split}
\end{equation}
\end{widetext}
leading to the full expression in  Eq.~\eqref{eq:comm2}.

Here, we have shown that arbitrary spin-adapted one-body and paired two-body operators can be expressed as a sum of operators 
acting between sequential spatial orbitals and their commutators.
Therefore, the product of exponential operators in the tUPS \textit{ansatz} with an increasing number of layers will eventually be able to
parametrise any exponential operation for the Lie algebra of many-body excitations, as indicated through successive applications of the Baker--Campbell--Hausdorff expansion
$\mathrm{e}^{\hat{A}} \mathrm{e}^{\hat{B}} = \mathrm{e}^{\hat{A}+\hat{B} + \frac{1}{2}[\hat{A},\hat{B}] + \cdots} $.\cite{GilmoreLieBook}
Consequently, exact wave functions within  a finite basis set can be obtained with a sufficient $L$, which we expect to be finite in practice.

\section{Orbital optimisation}
\label{sec:oo}
Orbital optimisation is essential to obtain smooth potential energy surfaces with truncated tUPS approximations.
The molecular spatial orbitals $\ket{\phi_p}$ are defined as a linear combination of $N$ atomic orbitals $\ket{\chi_\mu}$ as 
\begin{equation}
\ket{\phi_p} = \sum_{\mu}^N \ket{\chi_\mu} C^{\mu \cdot}_{\cdot p}, 
\end{equation}
where we use the  nonorthogonal tensor notation.\cite{HeadGordon1998}
For real orbitals, the coefficients $C^{\mu \cdot}_{\cdot p}$ are orthogonalised as
\begin{equation}
\sum_{\mu\nu}^N C^{\cdot \mu}_{ p \cdot} \braket{\chi_\mu}{\chi_\nu}  C^{\nu \cdot}_{\cdot q} = \delta_{pq}. 
\end{equation}
Variations in the orbitals for the correlated wave function $\ket{\Psi}$ can be parametrised
using the exponential form
\begin{equation}
\ket{\Psi(\bm{s})}  = \exp(\sum_{m > n}^N s_{mn} (\hE_{mn} - \hE_{nm})) \ket{\Psi},
\end{equation}
where $\bm{s}$ is the lower triangle of an $N \times N$  anti-Hermitian matrix representing a step in the orbital space.
The energy can be optimised with respect to the orbital coefficients by computing the gradient in this 
 representation as\cite{HelgakerBook}
\begin{equation}
\left.\frac{\partial E}{\partial s_{mn}}\right\vert_{\bm{s}=\bm{0}} = 2 (F_{mn} - F_{nm}),
\end{equation}
where $F_{mn}$ are matrix elements of the generalized Fock matrix and are defined as
\begin{equation}
F_{mn} = \sum_q^N D_{mq} h_{nq} + \sum_{qrs}^N \Gamma_{mqrs} g_{nqrs}.
\end{equation}
Here, $D_{pq} = \mel{\Psi}{\hat{E}_{pq}}{\Psi}$ and $\Gamma_{pqrs} = \mel{\Psi}{\hat{e}_{pqrs}}{\Psi}$ are the 
one- and two-body reduced density matrices (RDMs) for the correlated wave function, respectively, and 
$\hat{e}_{pqrs} = \hat{E}_{pq} \hat{E}_{rs} - \delta_{qr} \hat{E}_{ps}$.\cite{HelgakerBook}
The one-electron and two-electron integrals are defined as $h_{pq} = \mel{\phi_p}{\hat{h}}{\phi_q}$ and
$g_{pqrs} = (pq|rs)$, respectively.
The optimal step can then be implemented by re-computing 
the one- and two-electron integrals using the updated orbitals 
\begin{equation}
\tilde{\bm{C}} = \bm{C} \exp(\bm{s}).
\end{equation}
Since this process only needs the one- and two-body RDMs, which are already measured to compute
the energy on a quantum device, this orbital optimisation does not require any additional quantum resources.

\section{Connection to perfect pairing theory}
\label{sec:pp}

Perfect pairing (PP) theory is a well-established classical valence bond approximation  for molecular systems.\cite{Lawler2010, Beran2005,Voorhis2000,Hurley1953,Lehtola2018}
Starting from a closed-shell Slater determinant, the PP wave function is constructed by assigning a unique virtual orbital $\phi_{i^*}$ 
to each occupied spatial orbital $\phi_i$ and considering all excitations between each occupied-virtual orbital pair. 
Arbitrary mixing between the singlet states for each occupied-virtual orbital pair can be parametrised with two rotation angles $(x_i, y_i)$
and the total wave function for $n$ electron pairs can then be expressed as the anti-symmetrised product
\begin{equation}
\begin{split}
\ket{\Psi_{\text{PP}}} = \hat{\mathcal{A}} \prod_{i=1}^n \Big( 
&\cos x_i  \cos y_i\, \phi_{i} \phi_{\bar{i}} + \sin y_i\, \phi_{i^*} \phi_{\bar{i}^*} 
\\ &+ \frac{ \sin x_i  \cos y_i}{\sqrt{2}}\, \qty(\phi_{i} \phi_{\bar{i}^*} + \phi_{i^*}\phi_{\bar{i}})  \Big) .
\end{split}
\label{eq:pp}
\end{equation}
This wave function corresponds to a product of non-interacting singlet-coupled electron pairs and has been instrumental 
in developing our understanding of molecular bonding.\cite{Hurley1953,Dykstra1980}
It can accurately describe the dissociation of a single bond, but requires additional improvements to capture strongly interacting pairs 
of spin-coupled electrons.\cite{Parkhill2010,Small2009}
The equivalence between the first half-layer of the pp-tUPS \textit{ansatz} (see Fig.~2) and the PP wave function arises because
the operators $\hat{U}_{2p,2p-1}^{(1)}$ can exactly solve each (2e, 2o) sub-problem, giving a direct product of exact two-orbital sub-problems analogous
to Eq.~\eqref{eq:pp}.

\bibliographystyle{Science}
\bibliography{master}

%
%
%

%

\end{document}


\title{Accurate and gate-efficient quantum ans\"atze for electronic states without adaptive optimisation: Supplemental Material}

\author{Hugh~G.~A.~Burton}
\email{hgaburton@gmail.com}

\date{\today}

\maketitle
\raggedbottom

\onecolumngrid

\vspace{-3em}
\section{Molecular geometries}
All nuclear coordinates are represented in angstroms.
The bond distance for linear molecules is provided in the main text.

\subsection{Triangular \ce{H6}}

Regular lattice with internuclear distance $R(\ce{H-H}) = \SI{2.0}{\angstrom}$.

\begin{verbatim}
H   0.0000   0.0000   0.0000
H   1.0000   1.7321   0.0000
H   2.0000   0.0000   0.0000
H   3.0000   1.7321   0.0000
H   4.0000   0.0000   0.0000
H   2.0000   3.4641   0.0000
\end{verbatim}

\subsection{Benzene}

Experimental structure obtained from the National Institute of Standards and Technology database.\cite{NIST}

\begin{verbatim}
C   1.3970   0.0000   0.0000
C   0.6985   1.2098   0.0000
C  -0.6985   1.2098   0.0000
C  -1.3970   0.0000   0.0000
C  -0.6985  -1.2098   0.0000
C   0.6985  -1.2098   0.0000
H   2.4810   0.0000   0.0000
H   1.2405   2.1486   0.0000
H  -1.2405   2.1486   0.0000
H  -2.4810   0.0000   0.0000
H  -1.2405  -2.1486   0.0000
H   1.2405  -2.1486   0.0000
\end{verbatim}

\subsection{Water}

\begin{itemize}[itemsep=-0.1em]
\item Variable  bond length $R(\ce{O-H})$.
\item Fixed bond angle $\angle\ce{HOH} = \SI{104.5}{\deg}$.
\end{itemize}

\subsection{\ce{Be  + H2} insertion}

A reaction pathway for the insertion of \ce{Be} into \ce{H2} was constructed following  
the strategy of Refs.~\cite{Evangelista2011} and \cite{Magoulas2023a}.
Keeping the \ce{Be} atom fixed, the \ce{H} nuclear coordinates were varied as
\begin{verbatim}
Be   0.00000  0.00000  0.00000
H    x       -y        0.00000
H    x        y        0.00000
\end{verbatim}
using a grid defined as $x_n = \frac{5.746708}{29}n$ and $y_n = 1.31011 - 0.164183\,x_n$ for $n=0,1,\dots,29$.

\subsection{Methylene}
\begin{itemize}[itemsep=-0.1em]
\item Fixed bond length $R(\ce{C-H}) =  \SI{1.117}{\angstrom}$.
\item Variable bond angle $\angle\ce{HCH}$.
\end{itemize}

\subsection{Tetramethyleneethane}

Structure obtained from \textit{ab initio} geometry optimisation using restricted Hartree--Fock (STO-3G).

\begin{verbatim}
C   1.5191359612  -1.2435560240   0.0000000000
C   0.7596227794   0.0236969349  -0.0000000000
C  -0.7595397676  -0.0236632811  -0.0000000000
C   1.4388758359   1.1618268649  -0.0000000000
C  -1.4388154450  -1.1617811361  -0.0000000000
C  -1.5190270294   1.2436044768   0.0000000000
H   1.0485608349  -2.2118574675   0.0000000000
H   2.5977393829  -1.2079530524   0.0000000000
H   2.5193775927   1.1619199006  -0.0000000000
H   0.9765986939   2.1353834069  -0.0000000000
H  -0.9765601118  -2.1353473098  -0.0000000000
H  -2.5193178879  -1.1618520565  -0.0000000000
H  -2.5976311650   1.2080244996   0.0000000000
H  -1.0484295180   2.2118956990   0.0000000000
\end{verbatim}

\bibliographystyle{Science}
\bibliography{master}